\documentclass[aps,pra,reprint,superscriptaddress,amsmath,amssymb,floatfix]{revtex4-1}
\usepackage{epsfig}
\usepackage[colorlinks=true,citecolor=blue,linkcolor=blue,urlcolor=blue]{hyperref}
\usepackage{siunitx}
\usepackage{nicefrac}
\usepackage{tabularx}
\usepackage{apjfonts}
\begin{document} 

\title{New Measurement Resolves Key Astrophysical {Fe}~{\normalfont\textsc{xvii}} Oscillator Strength Problem}

\author{Steffen~ K\"uhn}\email[Corresponding author.\\ ]{steffen.kuehn@mpi-hd.mpg.de}
\affiliation{Max-Planck-Institut f\"ur Kernphysik, Saupfercheckweg 1, 69117 Heidelberg, Germany}%
\affiliation{Heidelberg Graduate School of Fundamental Physics, Ruprecht-Karls-Universit\"at Heidelberg, Im Neuenheimer Feld 226, 69120 Heidelberg, Germany}

\author{Charles~Cheung}
\affiliation{Department of Physics and Astronomy, University of Delaware, Newark, Delaware 19716, USA}

\author{Natalia~S.~Oreshkina}
\affiliation{Max-Planck-Institut f\"ur Kernphysik, Saupfercheckweg 1, 69117 Heidelberg, Germany}

\author{Ren\'e~Steinbr\"ugge}
\affiliation{Deutsches Elektronen-Sychrotron DESY, Notkestra{\ss}e 85, 22607 Hamburg, Germany}%

\author{Moto~Togawa}%
\affiliation{Max-Planck-Institut f\"ur Kernphysik, Saupfercheckweg 1, 69117 Heidelberg, Germany}

\author{Sonja~Bernitt}
\affiliation{Max-Planck-Institut f\"ur Kernphysik, Saupfercheckweg 1, 69117 Heidelberg, Germany}%
\affiliation{Institut f\"ur Optik und Quantenelektronik, Friedrich-Schiller-Universit\"at Jena,  Max-Wien-Platz 1, 07743 Jena, Germany}%
\affiliation{Helmholtz-Institut Jena, Fr\"obelstieg 3, 07743 Jena, Germany}%
\affiliation{GSI Helmholtzzentrum f\"ur Schwerionenforschung, Planckstra{\ss}e 1, 64291 Darmstadt, Germany}%

\author{Lukas~Berger}
\affiliation{Max-Planck-Institut f\"ur Kernphysik, Saupfercheckweg 1, 69117 Heidelberg, Germany}%

\author{Jens~Buck}
\affiliation{Institut f\"ur Experimentelle und Angewandte Physik (IEAP), Christian-Albrechts-Universit\"at zu Kiel, Leibnizstr. 11-19, 24098 Kiel, Germany}

\author{Moritz~Hoesch}
\affiliation{Deutsches Elektronen-Sychrotron DESY, Notkestra{\ss}e 85, 22607 Hamburg, Germany}%

\author{J\"orn~Seltmann}
\affiliation{Deutsches Elektronen-Sychrotron DESY, Notkestra{\ss}e 85, 22607 Hamburg, Germany}%

\author{Florian~Trinter}
\affiliation{Institut f\"ur Kernphysik, Goethe-Universit\"at Frankfurt am Main, Max-von-Laue-Stra{\ss}e 1, 60438 Frankfurt am Main, Germany}
\affiliation{Molecular Physics, Fritz-Haber-Institut der Max-Planck-Gesellschaft, Faradayweg 4-6, 14195 Berlin, Germany}

\author{Christoph~H.~Keitel}
\affiliation{Max-Planck-Institut f\"ur Kernphysik, Saupfercheckweg 1, 69117 Heidelberg, Germany}%

\author{Mikhail~G.~Kozlov}
\affiliation{St.\ Petersburg Electrotechnical University ``LETI'', Prof.\ Popov Str.\ 5, St.\ Petersburg, 197376, Russia}

\author{Sergey~G.~Porsev}
\affiliation{Department of Physics and Astronomy, University of Delaware, Newark, Delaware 19716, USA}

\author{Ming~Feng~Gu}
\affiliation{Space Science Laboratory, University of California, Berkeley, CA 94720, USA}%

\author{F.~Scott~Porter}
\affiliation{NASA/Goddard Space Flight Center, 8800 Greenbelt Rd, Greenbelt, MD 20771, USA}%

\author{Thomas~Pfeifer}
\affiliation{Max-Planck-Institut f\"ur Kernphysik, Saupfercheckweg 1, 69117 Heidelberg, Germany}%
   
\author{Maurice~A.~Leutenegger}
\affiliation{NASA/Goddard Space Flight Center, 8800 Greenbelt Rd, Greenbelt, MD 20771, USA}%

\author{Zolt\'an~Harman}
\affiliation{Max-Planck-Institut f\"ur Kernphysik, Saupfercheckweg 1, 69117 Heidelberg, Germany}%

\author{Marianna~S.~Safronova}
\affiliation{Department of Physics and Astronomy, University of Delaware, Newark, Delaware 19716, USA}
 
\author{Jos\'e~R.~Crespo~L\'opez-Urrutia}%
\affiliation{Max-Planck-Institut f\"ur Kernphysik, Saupfercheckweg 1, 69117 Heidelberg, Germany}%

\author{Chintan~Shah}\email[Corresponding author.\\]{chintan.shah@mpi-hd.mpg.de}
\affiliation{NASA/Goddard Space Flight Center, 8800 Greenbelt Rd, Greenbelt, MD 20771, USA}%
\affiliation{Max-Planck-Institut f\"ur Kernphysik, Saupfercheckweg 1, 69117 Heidelberg, Germany}%

\vspace{4mm}

\date{Received 22 January 2022; revised 27 July 2022; accepted 23 August 2022; published 5 December 2022}

\vspace{2mm}

\begin{abstract}
One of the most enduring and intensively studied problems of X-ray astronomy is the disagreement of state-of-the art theory and observations for the intensity ratio of two Fe{~\sc xvii} transitions of crucial value for plasma diagnostics, dubbed 3C and 3D. 
We unravel this conundrum at the PETRA~III synchrotron facility by increasing the resolving power 2.5 times and the signal-to-noise ratio thousandfold compared with our previous work. 
The Lorentzian wings had hitherto been indistinguishable from the background and were thus not modeled, resulting in a biased line-strength estimation. 
The present experimental oscillator-strength ratio $R_\mathrm{exp}=f_{\mathrm{3C}}/f_{\mathrm{3D}}=3.51(2)_{\mathrm{stat}}(7)_{\mathrm{sys}}$ agrees with our state-of-the-art calculation of $R_\mathrm{th}=3.55(2)$, as well as with some previous theoretical predictions. 
To further rule out any uncertainties associated with the measured ratio, we also determined the individual natural linewidths and oscillator strengths of 3C and 3D transitions, which also agree well with the theory. 
This finally resolves the decades-old mystery of Fe{~\sc xvii} oscillator strengths.
\end{abstract}

\maketitle

%
%

Soft X-ray spectra from present space observatories such as {\it Chandra} \cite{weisskopf2000chandra} and {\it XMM-Newton} \cite{jansen2001xmm,den2001reflection} offer deep insight into massively energetic astrophysical sources, and together with next-generation future missions such as {\it XRISM} \cite{xrism2018} and {\it Athena} \cite{barret2016,pajot2018athena} will continue improving our understanding of phenomena driving the growth of galaxies, star formation, and the reionization phase of the universe. 
Since ubiquitous iron ions are among the strongest emitters in most X-ray sources, decades of efforts have been aimed at understanding their spectra, and delivering valuable diagnostics of electron temperature, density, chemical abundances, and opacities of various astrophysical objects~\cite{parkinson1973new,smith1985,schmelz1992,waljeski1994,phillips1996,behar2001chandra,gu2003FeL,pfk2003,beiersdorfer2018,gu2019,gu2020capellaEBIT,grell2021}. 
One of the key diagnostics uses the ratio of two strong emission lines from the same ion which have different oscillator strengths to probe for resonance scattering in an otherwise optically thin plasma. This provides constraints on gas turbulence and column density in stellar coronae, as well as in diffuse hot gas in galaxy clusters and groups, and their lower mass analogs, giant elliptical galaxies~\cite{xu2002high, werner2009constraints, pzw2012, hitomi2016, ogorzalek2017improved}. 

The L-shell emission of Ne-like Fe{~\sc xvii} (Fe$^{16+}$), namely the resonance line 3C ($[2p^6]_{J=0} \rightarrow [(2p^5)_{1/2}\,3d_{3/2}]_{J=1}$) and the intercombination line 3D ($[2p^6]_{J=0} \rightarrow [(2p^5)_{3/2}\,3d_{5/2}]_{J=1}$), are often the best candidates for such diagnostics in plasmas with temperatures of 1--10 MK, given their brightness. 
However, their intensity ratio has long been the subject of controversy~\cite{bernitt2012unexpectedly}.
Early solar observations gave a ratio of 3C/3D below 2.7, whereas atomic theory and astrophysical models predicted a ratio close to 4.0~\cite{blake1965spectral,parkinson1973new,mckenzie1980solar,smith1985,loulergue1975study,waljeski1994,bhatia1992,cornille1994radiative,saba1999,brown1998laboratory}. 
This difference was initially attributed to resonance scattering in the Sun~\cite{schmelz1992,phillips1996}. However, systematic measurements using an electron beam ion trap (EBIT)~\cite{levine1988electron,brown2001diagnostic} showed significant blending of inner-shell Fe{~\sc XVI} line C ($[2p^6 3s]_{J={1/2}}\rightarrow[(2p^5)_{1/2} (3s 3d)_{5/2}]_{J={3/2}}$) with line 3D, which could potentially reduce the apparent 3C/3D ratio and, if not accounted for, lead to an overestimate of the strength of resonance scattering~\cite{brickhouse2005}. 
Theoretical predictions also improved over time with the gradual inclusion of more configurations and relativistic effects, converging close to a ratio of $\approx3.5$~\cite{zhang1989relativistic,cornille1994radiative,kaastra2011high,dong2003theoretical,loch2005effects,chen2007converged,gu2009new,jonsson2014,wu2021angular}, but still exceeding measurements by $\sim$10--15\%~\cite{brown1998laboratory,laming2000emission,beiersdorfer2001,beiersdorfer2002laboratory,beiersdorfer2004laboratory,brown2006energy,gillaspy2011fe,shah2019revisiting}.
This has compromised the diagnostic utility of Fe{~\sc xvii} lines, and repeatedly raised questions regarding the accuracy of calculations and models. 

\begin{figure*}[t]
	\includegraphics[width=0.9\textwidth]{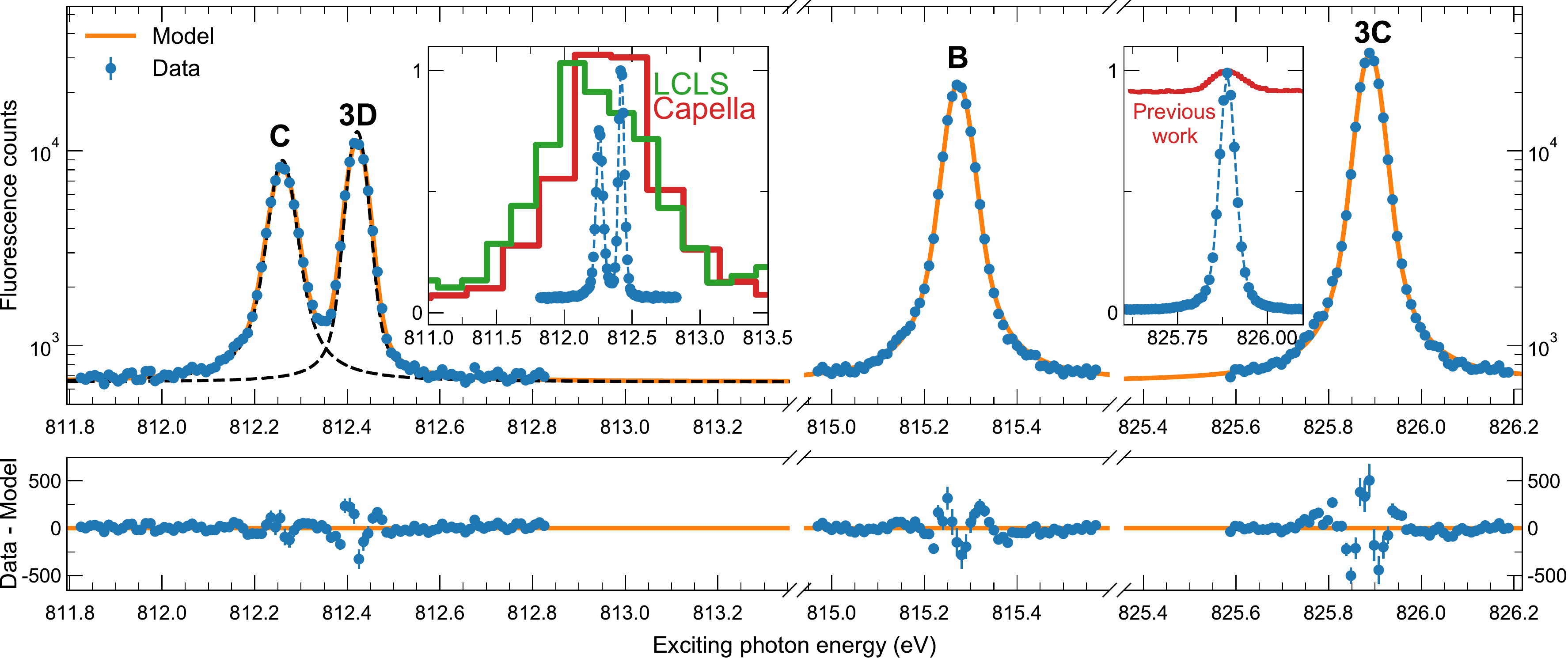}
	\caption{
		Fluorescence yield of the soft X-ray transitions 3C and 3D of Fe{~\sc xvii} as well as B and C of Fe{~\sc XVI} as a function of the exciting-photon energy. Fitted Voigt profiles (orange) show residuals (bottom panel) due to a nonperfectly Gaussian distribution of the monochromator spectral bandwidth. Inset (right): 3C comparison between present and previous~\cite{kuhn2020high} works. Inset (left): resolved C and 3D from this work compared to unresolved in LCLS measurements (green~\cite{bernitt2012unexpectedly}) and Capella observations (red~\cite{huenemoerder2011tgcat}).  
	}
	\label{fig:3C3DSpectra}
\end{figure*}

The line strengths of 3C and 3D in a collisionally excited plasma are expected to scale with their oscillator strengths, with only small contributions from resonant excitation. 
Further complications in astrophysical plasmas and experiments can arise from neglect of dielectronic recombination, charge exchange, inner-shell ionization, collisional quenching, polarization, and non-Maxwellian electron-energy distribution.
Thus, an experiment using only photons to populate the upper levels of 3C and 3D has the advantage of testing and validating only atomic structure and avoiding more complex collision physics or other plasma processes.
A resonant photoexcitation experiment at the free-electron laser Linac Coherent Light Source (LCLS) measured an oscillator-strength ratio $R= f_{\mathrm{3C}}/f_{\mathrm{3D}} = 2.61\pm0.21$, which unexpectedly departed from theory even more than earlier works measuring collision strengths~\cite{bernitt2012unexpectedly}. 
Nonlinear excitation dynamics~\cite{ock2014,oreshkina2016x,Loch_2015,fogle2017} and charge-state population transfer \cite{wu2019change} were suggested as sources of this surprising discrepancy. 
Renewed experimental efforts with a synchrotron-based technique brought tenfold resolution improvement compared with the LCLS experiment and furthermore excluded population transfer and nonequilibrium effects because of their much lower instantaneous photon flux. 
The result $R=3.09\pm0.1$~\cite{kuhn2020high} still disagreed with the greatly improved, large-scale converged calculations from the same work, which took into account all known quantum mechanical effects, and predicted $R=3.55 \pm 0.05$~\cite{kuhn2020high}.

In this Letter, we present the most precise measurements of the oscillator-strength ratio of resonantly photoexcited 3C and 3D lines in Fe{~\sc xvii} ions, with an increase in both the resolving power by a factor of 2.5 and the signal-to-noise ratio (SNR) by 3 orders of magnitude compared with our previous work~\cite{kuhn2020high}.
Both these improvements allowed us to determine the individual natural linewidths of 3C and 3D. 
Since natural line broadening is directly proportional to the oscillator strengths~\cite{radpropbook}, the complexities arising from the various upper-level population mechanisms, perhaps the most intractable difficulty in previous measurements, can now be eliminated. 
Our improved large-scale atomic calculations also have reduced uncertainty by a factor of 2.5 compared to~\cite{kuhn2020high}.
Finally, the present experiment agrees with our present and previous predictions and resolves the decades-old Fe{~\sc xvii} oscillator-strength paradox.

%
%

\begin{figure*}[t]
\includegraphics[width=0.9\textwidth]{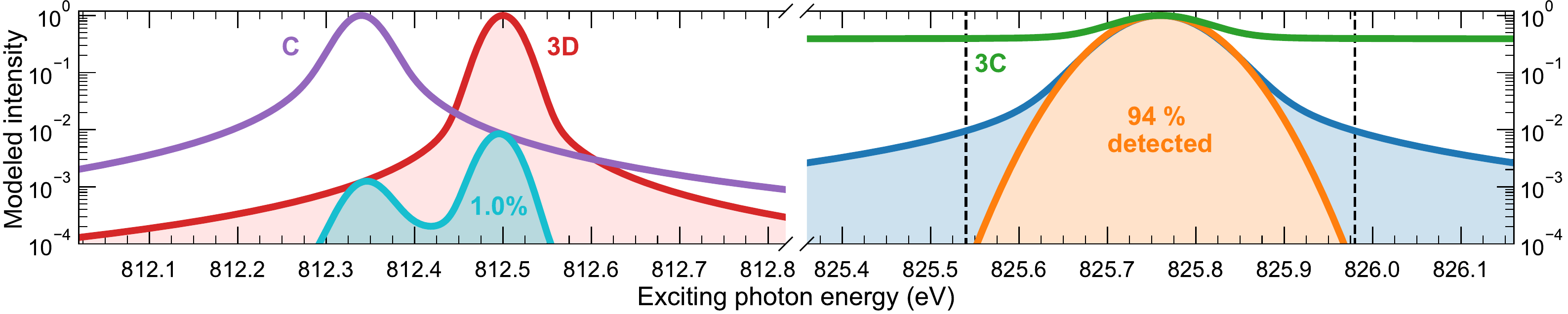}
\caption{Left: Synthetic Voigt profiles for C (purple) and 3D (red). Predicted natural linewidths are convolved with a Gaussian component. At similar line intensities, the overlapping product is only 1\% of the total 3D intensity, which suppresses feeding of the 3D lower state in Fe{~\sc xvii} from the upper level of C in Fe{~\sc xvi} through an autoionization decay branch of $\approx 60$\%. Right: Synthetic 3C Voigt profile without (blue) and with a strong background (green) as in~\cite{kuhn2020high} for comparison. A Gaussian model (orange) fitted to this curve can hardly separate the Lorentzian wings from the background and recovers only 94\% of the true intensity. There, a Voigt fit to the data from~\cite{kuhn2020high} also does not work well at the wings due to the low SNR.}
\label{fig:amplitudes}
\end{figure*}

The experiment was performed using the PolarX-EBIT~\cite{micke2018} at the PETRA~III beamline P04~\cite{viefhaus2013variable}. 
PolarX-EBIT generates a nearly monoenergetic electron beam---using a novel off axis gun designed for photonic studies of highly charged ions (HCI)~\cite{epp2007soft,simon2010b,rudolph2013x,steinbrugge2015absolute,Leutenegger2020,Togawa2020}---which is compressed to a radius of 50~$\mu$m by a 0.87 T magnetic field. 
The HCIs are produced by successive electron impact ionization, and are simultaneously trapped in the radial direction by the negative space-charge potential of the electron beam and in the axial direction by applied potentials on the drift tubes.
The electron impact ionization cross sections peak at energies 2 to 3 times higher than the threshold. Thus, we set the beam energy $E_{\mathrm{B}} \approx \SI{1200}{\electronvolt}$, which is approximately 2.5 times the production threshold of Fe{~\sc xvii}, 490 eV. 
However, this beam energy exceeds the excitation thresholds of 3C and 3D, causing a very strong and constant X-ray background induced by direct electron-impact excitation~\cite{brown2006energy,shah2019revisiting}. 
At free-electron lasers (FELs), such high background can be suppressed by using time coincidence between the photon-excited fluorescence signal and the pulsed photon train~\cite{epp2007soft,epp2010,bernitt2012unexpectedly}. 
However, this was not possible here because synchrotrons exhibit a much shorter photon pulse interval compared with FELs and due to the limited time resolution of silicon drift detectors (SDDs). 
Indeed, this was the main reason behind a poor SNR of only $\approx 0.05$ in \cite{kuhn2020high}.

To overcome this difficulty, we reduced the strong electron-impact induced background by cyclically varying $E_{\mathrm{B}}$ within tens of nanoseconds between a breeding and probing phase, using fast switching power supplies (see Fig.~2 in the Supplemental Material~\cite{SuppMat}~\nocite{beiersdorfer2000x,mackel2011laser,henke1993x,AK,wang2017comment,kahl2019ambit}). 
First, we produced a sufficient amount of Fe{~\sc xvii} at breeding energy of $E_{\mathrm{B}} = \SI{1200}{\electronvolt}$ for 200~ms.
Then, we reduced $E_{\mathrm{B}}$ to the probing energy of $\SI{250}{\electronvolt}$ within a few tens of ns.
The advantage of such low probing energy is that we can eliminate the background from dielectronic recombination, resonant excitation, and direct electron-impact excitation channels ({cf.} Fig. 1 of~\cite{shah2019revisiting}). 
Moreover, the ion population remained mostly unchanged within the switching time because the collision rates are relatively low at electron densities below $10^{10}~\mathrm{cm}^{-3}$, typically found in PolarX-EBIT~\cite{grilo2021}.
On the other hand, the probing energy cannot further produce Fe{~\sc xvii}, leading to a continuous depletion of Fe{~\sc xvii} due to radiative recombination and charge exchange~\cite{grilo2021}, as well as due to loss of ions from the trap~\cite{penetrante1991}. 
We observed a confinement half-life of $\sim50$ ms for Fe{~\sc xvii}. 
Once the Fe{~\sc xvii} population is completely depleted, the cycle starts again to produce ions at the breeding energy.
By optimizing both the duty cycle and the slew rate for a maximum resonant photoexcitation signal, we achieved a SNR $\approx45$, approximately a thousandfold improvement compared with our previous work~\cite{kuhn2020high}.

The monochromator spectral resolution increases linearly with decreasing aperture size up to a limit set by the source size.
Thus, with a significant improvement in SNR, we were also able to decrease the monochromator exit-slit width from \SI{50}{\micro\meter} (as in~\cite{kuhn2020high}) to \SI{25}{\micro\meter} to improve the spectral resolving power while maintaining the strong resonant photoexcitation signal for lines 3C and 3D. 
This in turn allowed us to fine tune other X-ray optical elements of the beamline, resulting in a further increase in spectral resolution~\cite{viefhaus2013variable, follath}. 
All these optimizations resulted in an excellent resolving power of $E/\Delta E=20\,000$ at 830 eV photon energy.

%
%

Both these unprecedented advances in SNR and spectral resolution are essential for accurately determining the line intensities and their profiles required to extract the natural linewidth of a given transition. 
By scanning the monochromator energy, we measured 3C and 3D of Fe{~\sc xvii} along with two other lines of Fe{~\sc xvi}, named B ($[2p^6 3s]_{J={1/2}}\rightarrow[(2p^5)_{1/2} (3s 3d)_{3/2}]_{J={1/2}}$) and C. 
Fluorescence X-rays from the decay of photoexcited ions were detected using two identical SDDs mounted side-on to the photon beam axis.
As shown in Fig.~\ref{fig:3C3DSpectra}, the statistical data quality allowed us to reliably fit each transition with a Voigt function, which is the convolution of Lorentz and Gaussian functions. 
The Gaussian part of the line shape stems from the thermal motion of the trapped ions, so-called Doppler broadening, and from the limited monochromator instrument resolution, while, in principle, the Lorentzian part of the line shape should only result from the natural linewidth of an electronic transition due to its finite intrinsic lifetime. 
We derived the area under the peak by fitting the Voigt function to lines 3C, 3D, C, and B, each scanned over 20 times. 
The weighted ratio of 3C flux to that of 3D, which is proportional to the oscillator-strength ratio, is $R = 3.51(2)$. 

Our own recent work~\cite{kuhn2020high} clearly resolved C from 3D. However, fits considering only a Gaussian model underestimated the Lorentzian line wings, which were buried in the background~\cite{laming2000emission}, and resulted in a biased line flux determination, as shown in Fig.~\ref{fig:amplitudes}. 
Now, with blends resolved, with perfectly exposed Lorentzian wings, and with continuously monitoring of background and photon-flux variations using side-on SDDs and a downstream calibrated photon diode, respectively, we could fit Voigt profiles with relative intensity uncertainties below $1\%$, achieving a statistical accuracy in $R$ that supersedes all previously reported experimental results.

Armed with this high resolution, we found in a later campaign at beamline P04 on a different ion an additional source of uncertainty.
While linearly scanning the monochromator for fluorescence, we let the photon beam passing through the EBIT reach an electron spectrometer with very high resolution ({\sc asphere}~\cite{ROSSNAGEL2001}) for simultaneously monitoring the excess energy of copious photoelectrons emitted from a clean gold surface. 
Periodic excursions of their energy from the expected values revealed that the angular encoder for the grating was inaccurately interpolating readings between fixed 0.1\,degree marks~\cite{Krempasky2011}. This causes an apparent cyclic compression and stretching of the photon-energy increments by $\approx\pm 0.04$\,eV with $\approx1$\,eV period (see Supplemental Material \cite{SuppMat}). 
We modeled the effects of this distortion on the Voigt profiles, resulting in a systematic uncertainty of $\approx$2\% that now dominates the final error budget of the 3C/3D ratio.
The final experimental result is $R=3.51(2)_{\mathrm{stat}} (7)_{\mathrm{sys}}$.

Given the observed line shape at an excellent resolution, the individual natural linewidths of 3C and 3D can also be determined. 
In general, the oscillator-strength ratio of two given electronic transitions, as in the case of 3C and 3D, is correlated with the corresponding ratio of natural linewidths through the following relation,
\begin{equation}
\frac{f_{\mathrm{3C}}}{f_{\mathrm{3D}}}  = \left(\frac{\Gamma_{\mathrm{3C}}}{\Gamma_{\mathrm{3D}}}\right) \Big/ \left(\frac{E_{\mathrm{{3C}}}}{E_{\mathrm{{3D}}}}\right)^2.
\label{ref:eq1}
\end{equation}
Hence, by combining the observed $R=3.51\pm0.09$ with the energy quotient $\left(E_{\mathrm{3C}}/E_{\mathrm{3D}}\right)^2 = 1.032$ determined in \cite{brown1998laboratory}, a linewidth ratio $\Gamma_{\mathrm{3C}} / \Gamma_{\mathrm{3D}}$ of $\approx$3.6 is expected. 
In contrast, the present ratio of the inferred Lorentzian widths of 3C and 3D lies well below 2.0. 
We have found the reason behind this by measuring the Lorentzian widths of well-known He-like F~{\sc viii} $K_\alpha, K_\beta$, and $K_\gamma$ transitions, for which the theory is deemed accurate. 
This revealed a pseudo-Lorentzian instrumental contribution of $7.0 \pm 0.3$ meV, which artificially increases the apparent natural linewidth of all observed transitions. 
It results from X-ray diffraction from elements of the beamline and mimics a Lorentzian profile at finite resolution~\cite{follath}. 
This effect was found to be nearly constant within the photon-energy range of~$E_{\mathrm{3C}}-E_{\mathrm{3D}}\approx 13.4$\,eV.
Furthermore, the convolution of two Lorentzians results in another Lorentzian whose width is simply the sum of the two widths.
Thus, we can accurately determine the difference $\Delta\Gamma_{\mathrm{3C}-\mathrm{3D}} = \SI{10.92\pm1.75}{}$\,meV. 
Now by combining this difference with Eq.~\ref{ref:eq1}, we can extract the individual natural linewidths of 3C and 3D as,
\begin{equation}
	\Gamma_{\mathrm{3C}} =
	\frac{\Delta\Gamma_{\mathrm{{3C-3D}}}}{1-{\frac{f(\mathrm{{3D}})}{f(\mathrm{{3C}})}
			\,\, \left(\frac{E_{\mathrm{{3D}}}}{E_{\mathrm{{3C}}}}\right)^2}},
	\mathrm{and\,\,\,}
	\Gamma_{\mathrm{3D}} =
	\frac{\Delta\Gamma_{\mathrm{{3C-3D}}}}{\frac{f(\mathrm{{3C}})}{f(\mathrm{{3D}})}
		\,\, \left(\frac{E_{\mathrm{{3C}}}}{E_{\mathrm{{3D}}}}\right)^2 - 1}.
	\label{3c_3d_gamma}
\end{equation}
Analogously, the natural linewidths of lines B and C are determined by adding their respective difference from 3D Lorentzian width to the inferred 3D natural linewidth. 
In a second approach, we used the extracted pseudo-Lorentzian beamline component from measured narrow transitions of {F~\sc viii} to validate the values determined by Eq.~\ref{3c_3d_gamma}.   
Both independent analysis methods give consistent values, and they agree very well with the predictions (see Supplemental Material \cite{SuppMat}).  
The influence of the periodic energy shifts of the monochromator on linewidths was also modeled, giving a relative systematic uncertainty of $5\%$ in the final error budget, as shown in Tab.~\ref{table1}.

\begin{figure}[t]
	\includegraphics[width=0.9\columnwidth]{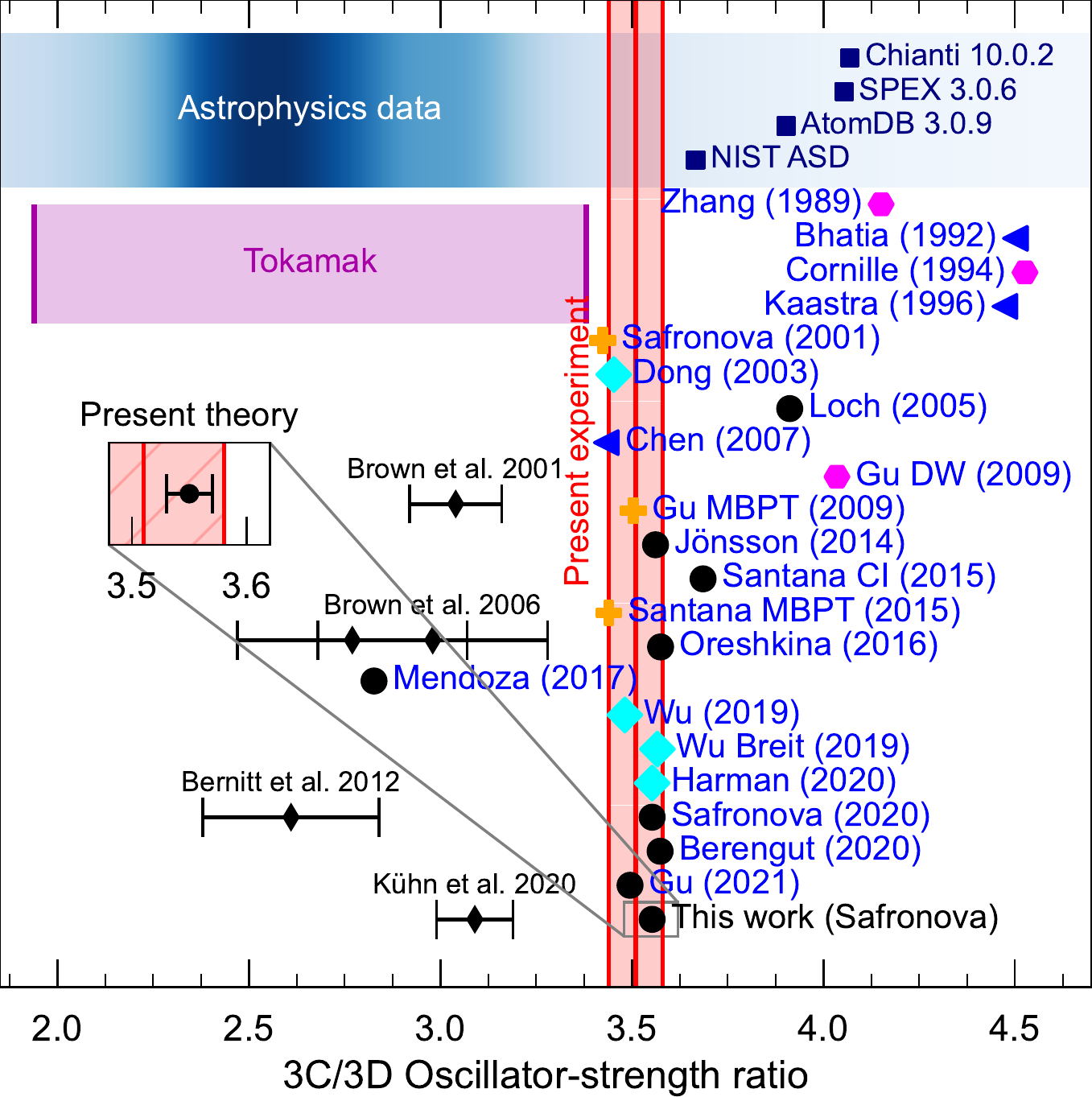}
	\caption{Present experimental 3C/3D oscillator-strength ratio compared with predictions, including our own, shown in an inset.
		Theoretical approaches \cite{zhang1989relativistic,bhatia1992,cornille1994radiative,kaastra1996spex,safronova2001electric,dong2003theoretical,loch2005effects,chen2007converged,gu2009new,jonsson2014,santana2015electron,oreshkina2016x,mendoza2017theoretical,wu2019change,kuhn2020high,gu2021,cheung2021scalable} are marked as follows: Distorted wave (magenta hexagons), multiconfiguration Dirac-Fock (teal diamonds), R-matrix (blue triangles), many-body perturbation theory (orange crosses), and configuration interaction (black circles). 
		Blue band: observed line ratios in astrophysical sources \cite{behar2001chandra,mewe2001chandra,xu2002high,blake1965,mckenzie1980solar,ness2003stellar}, with color shades coding the distribution of values weighted by their reported accuracies. 
		Purple band: spread of tokamak results~\cite{beiersdorfer2004laboratory}. 
		Black diamonds: previous EBIT results~\cite{brown2001diagnostic,brown2006energy,bernitt2012unexpectedly,kuhn2020high}. 
		Navy blue squares: spectral plasma models and databases~\cite{NIST_ASD,foster2012,kaastra1996spex,chianti2021}. 
		The data behind the figure are given in the Supplemental Material \cite{SuppMat}.}
	\label{fig:results}
\end{figure}

\begin{table}[b]
\centering
\caption{\label{table1} Average of two methods used to infer the natural linewidths ($\Gamma$ in meV) and oscillator strengths ($f$) compared with their predictions.}
\begin{tabular*}{\columnwidth}{c @{\extracolsep{\fill}} cccc}
\hline \hline
Line  & $\Gamma_\mathrm{exp}$ & $\Gamma_\mathrm{th}$ & $f_\mathrm{exp}$ & $f_\mathrm{th}$ \\
\hline
3C & \SI{ 15.07 \pm  1.07 }{} & \SI{ 14.74 \pm  0.01 }{} & \SI{  2.32 \pm  0.17 }{} & \SI{ 2.271 \pm 0.002 }{} \\
3D & \SI{  3.79 \pm  0.52 }{} & \SI{  4.03 \pm  0.02 }{} & \SI{  0.60 \pm  0.08 }{} & \SI{ 0.641 \pm 0.003 }{} \\
\hline
B & \SI{ 15.87 \pm  1.16 }{} & 14.43~\cite{oreshkina2016x} & \SI{  0.84 \pm  0.06 }{} & 0.760~\cite{oreshkina2016x} \\
C & \SI{ 20.07 \pm  1.38 }{} & 23.10~\cite{oreshkina2016x} & \SI{  2.13 \pm  0.15 }{} & 2.454~\cite{oreshkina2016x} \\
\hline \hline
\end{tabular*}%
\end{table}%

%
%

We also performed a large-scale configuration-interaction (CI) calculation using a new version of our highly scalable parallel CI code, where we substantially enlarged the basis compared with our previous calculations~\cite{kuhn2020high}. 
Details are given in the Supplemental Material~\cite{SuppMat}.
In brief, the new basis includes all orbitals up to $24spdfg12h$ compared with $12sp17dfg$ in~\cite{kuhn2020high}.
With 1.2 million configurations and almost 100 million Slater determinants, including triple excitations and opening of the $1s^2$ shell, we have fully saturated the CI expansion.
The Breit interaction~\cite{breit1929} and QED effects are included in these computations following Refs.~\cite{QED,kuhn2020high}.
QED contributes $-0.016$ to the ratio, mostly through its effect on the dipole matrix elements, whereas its influence on the 3C and 3D energies is negligible. 
Conservatively, we consider the total QED correction as the uncertainty of our final prediction of $R=3.55\pm0.02$. 
Similarly, the natural linewidths for 3C and 3D transitions are calculated to be $\SI{14.74\pm0.01}{\milli\electronvolt}$ and $\SI{4.03\pm0.01}{\milli\electronvolt}$, respectively. 
Theoretical $\Delta\Gamma_{\mathrm{3C}-\mathrm{3D}}$ is 10.71(2)\,meV, with the QED contribution of 0.02\,meV that is larger than the electronic correlation uncertainty. 
Overall, the QED calculation uncertainties, resulting from subtle differences in configuration mixing for the upper and lower states of 3C and 3D, dominate our predictions. 
%

%
%

Figure~\ref{fig:results} shows that the experimental oscillator-strength ratio and natural linewidths agree very well with our previous and present large-scale calculations (see Tab.~1 in the Supplemental Material \cite{SuppMat} for details). 
It is also noteworthy that certain earlier theoretical works agree with the present experimental results, while others disagree for various reasons, including differences in terms of number of configurations, incomplete description of correlation and relativistic effects, as well as insufficient numerical convergence.
On the contrary, previous EBIT and tokamak measurements~\cite{brown1998laboratory,laming2000emission,beiersdorfer2001,beiersdorfer2002laboratory,beiersdorfer2004laboratory,brown2006energy,gillaspy2011fe,shah2019revisiting}, and astrophysical observations show deviations from our measurement and from each other. 
At least some of the scatter in these results is attributable to known effects such as neglected blending of satellite lines~\cite{brown2001systematic,brickhouse2005}.
Nevertheless, a significant discrepancy remains between even the most careful measurements and calculations of collision strength ratios, which remains to be solved in future work~\cite{brown2006energy,shah2019revisiting}.

Our work exposes how critical a good understanding of non-Gaussian line shapes in both theory and experiment is for accurately determining transition amplitudes and line positions, especially when the resolving power is limited such that line wings are indistinguishable from background~\cite{PhysRevLett.127.235001}.
Our experimental technique achieved hitherto-unfeasible spectral resolution and SNR, both of which were crucial for allowing measurement of the Lorentzian component of line shapes and thus for accurately determining the natural linewidths of key astrophysical X-ray transitions.

Accurate oscillator strengths and natural linewidths are essential for modeling and computation of the mean opacity of hot-dense matter~\cite{fontes2015,pain2017,nagayama2019,PhysRevLett.127.235001}. 
The present experimental confirmation of state-of-the-art theories indicates that the oscillator strengths of Fe{~\sc xvii} cannot fully explain the discrepancy between measured and modeled iron opacity~\cite{bailey2007,bailey2015}, and thus the radiative transport in the interior of the Sun~\cite{iglesias2017}. 
In addition, the results shown in Fig.~\ref{fig:results} point to the urgent need for updating oscillator strengths in spectral databases and plasma modeling codes such as NIST~\cite{NIST_ASD}, AtomDB~\cite{smith2001,foster2012}, SPEX~\cite{kaastra1996spex}, and Chianti~\cite{chianti2021}, given the upcoming launch of XRISM~\cite{xrism2018} in early 2023, and of Athena~\cite{barret2016} in the 2030s. 
Scientific breakthroughs expected from these missions will only be realized with an improved understanding of atomic theory through targeted laboratory benchmarks~\cite{hitomi2018,heuer2021,gu2022}.

%
%

\hfill \break

\begin{acknowledgments}
Financial support was provided by the Max-Planck-Gesellschaft (MPG) and Bun\-des\-mi\-ni\-ste\-ri\-um f{\"u}r Bildung und Forschung (BMBF) through Project No.~05K13SJ2. 
C.S. acknowledges support by an appointment to the NASA Postdoctoral Program at the NASA Goddard Space Flight Center, administered by Oak Ridge Associated Universities under contract with NASA and by Max-Planck-Gesellschaft (MPG).
F.S.P. and M.A.L. acknowledge support from NASA's Astrophysics Program. 
The views and conclusions contained in this document are those of the authors and should not be interpreted as representing the official policies, either expressed or implied, of the National Aeronautics and Space Administration (NASA) or the U.S.Government. The U.S.Government is authorized to reproduce and distribute reprints for government purposes notwithstanding any copyright notation herein.
The theoretical research was supported in part by US NSF Grant No.\ PHY-2012068, European Research Council (ERC) under the European Union’s Horizon 2020 research and innovation program (Grant Agreement No. 856415), the Russian Science Foundation under Grant No.\ 19-12-00157, and through the use of Caviness and DARWIN HPC systems at the University of Delaware~\cite{Darwin}. 
We acknowledge DESY (Hamburg, Germany), a member of the Helmholtz Association HGF, for the provision of experimental facilities. Parts of this research were carried out at PETRA~III.
We thank Jens Viefhaus and Rolf Follath for valuable discussions on X-ray monochromator resolution and performance, and the synchrotron-operation team and P04 team at PETRA~III for their skillful and reliable work. 
We also thank anonymous referees whose comments and suggestions helped improve and clarify this manuscript.
\end{acknowledgments}

%
%
\bibliography{references}

\end{document}